\documentclass[letter, 12pt ,onecolumn]{IEEEtran}

\usepackage{times, epsfig}
\usepackage{amssymb}
\usepackage{amsmath}
\usepackage{graphicx}
\usepackage[latin1]{inputenc}
\usepackage{graphicx}
\usepackage{setspace}
\usepackage{psfrag}
\usepackage{cite}
\usepackage{multirow}
\usepackage{wasysym}
\usepackage{multirow}
\usepackage{float}
\usepackage{pdfsync}
\usepackage{pstricks}

\begin{document}

\title{The Quest for Sustainable Smart Grids}

\author{Pedro H. J. Nardelli\\Centre for Wireless Communications, University of Oulu}

\maketitle

The authors have discussed in \cite{Brummitt2013} the need for a transdisciplinary science to electric grids.
This letter aims at contributing to their view by commenting the so-called \textit{sustainable smart grids}.

Let us first define what we mean by \textit{sustainable}.
To be fair across its diverse definitions \cite{DeVries2012}, we borrow the Oxford dictionary meaning: ``able to be maintained at a certain rate or level''.
In this sense, we understand that the electric power grids should be sustainable - over time and space - in terms of not only technological aspects but also in other equally important dimensions such as social, economical, cultural and environmental.

This has been partly touched by \cite{Brummitt2013}.
However, how should one define the dimensions of this analysis without skipping from the real world operation?
Our answer also comes from complexity science: we believe that technological, social, economical, cultural and environmental processes co-evolve, shaping each other in time and space.
In other words, the (perceived) states of the system at a given time, locally and globally, are emerging properties of dialectical relations between the involved processes. 
The challenge is to develop theoretic models to characterize them in different contexts.

From this perspective, technological sciences urge new ways of looking at problems; transdisciplinary understanding needs to be built.
For example, engineers need to be aware of analytical sociology \cite{Hedstrom2010} to assess social processes and their underlying mechanisms.
They also need to see why economic sciences have historically failed to prevent big crises and why new theories are then needed \cite{Helbing2013}.

We could go on with other disciplines, but we prefer to turn the attention to the most important aspect of any scientific theory: \textit{validation}.
As clearly stated in \cite{Cockshott2011}: ``Math can be seductive(...) But when the maths claims to be a model of real world, beauty can mislead''.
Therefore, a scientific theory must be tested against real-world data to be valid (or not), otherwise it can be classified as ``not even wrong''.
Agent-based simulations, in combination with real data, are sometimes worth to validate theories \cite{DeVries2012,Hedstrom2010,Helbing2013}.

After this digression, let us come back to our problem. 
We have the following hypothesis: even though the technology is a necessary condition to the system, the smart grid will be sustainable if, and only if, all involved processes co-evolve in sustainable ways.
Hence the smart grid is context-dependent and one must consider social structures, economic regulations, environment, weather, acceptance of new technologies, land use, security etc. to have a proper picture of its evolution in different places.
We then expect sustainable smart grid models for Brazilian urban areas quite different from the ones for Finnish countryside.

All in all, constructing a successful theory for sustainable smart grids is a huge transdisciplinary quest, only possible when the involved processes, and their dialectical interrelations, are well-understood.
Otherwise, as already discussed by \cite{Brummitt2013} and reported in many places where some of the smart grid features have been implemented, such a technology might (and probably will) fail to deliver its promising benefits.

\section*{Acknowledgments}
The points arisen by this letter are in line with the activities carried out in the Brazilian Science without Borders Special Visiting Researcher fellowship CAPES n.076/2012, and the project \textit{A Theory for Sustainable Smart Grids: Combining Communication Theory, Power Systems, Signal Processing and Economics from a Complexity Science Perspective} under the Finnish Academy Sustainable Energy (SusEn) framework jointly funded by CNPq-Brazil n.490235/2012-3.

\end{document}